\documentclass[aps,floatfix]{revtex4}
\usepackage{graphicx}
\usepackage{epsfig}
\begin{document}

\title{Testing the LCDM model (and more) with the time evolution of the redshift}
\author{Kayll Lake$\;$\cite{email}}
\affiliation{Department of Physics, Queen's University, Kingston,
Ontario, Canada, K7L 3N6 }
\date{\today}
\begin{abstract}
With the many ambitious proposals afoot for new generations of
very large telescopes, along with spectrographs of unprecedented
resolution, there arises the real possibility that the time
evolution of the cosmological redshift may, in the not too distant
future, prove to be a useful tool rather than merely a theoretical
curiosity. Here I contrast this approach with the standard
cosmological procedure based on the luminosity (or any other
well-defined) distance. I then show that such observations would
not only provide a direct measure of all the associated
cosmological parameters of the LCDM model, but would also provide
wide-ranging internal consistency checks. Further, in a more
general context, I show that without introducing further time
derivatives of the redshift one could in fact map out the dark
energy equation of state should the LCDM model fail. A consideration
of brane-world scenarios and interacting dark energy models serves to emphasize the
fact that the usefulness of such observations would not be
restricted to high redshifts.
\end{abstract}
\maketitle
\section{Introduction}
Differentiation of the standard cosmological redshift
\cite{notation}
\begin{equation}\label{redshift}
 1+z=\frac{a(t_{o})}{a(t_{e})}=\frac{dt_{o}}{dt_{e}}
\end{equation}
gives the McVittie equation \cite{mcvittie}
\begin{equation}\label{mcvittie}
\dot{z}=H_{o}\left(1+z-\frac{H(z)}{H_{o}}\right) \equiv H_{o}Z(z),
\end{equation}
where the $t$ - derivative is evaluated today. The boundary
condition is $Z(0)=0$. The fact that $\dot{z}$ is very
difficult to detect \cite{detect} is due to the fact that $H_{o}$
is, in human terms, so very small ($ \sim h \times 10^{-10} \;yr^{-1}$
with $h \equiv H_{o}/100$ and $H_{o}$ in km s$^{-1}$ Mpc$^{-1}$
where, according to all observations, $h \sim 1$). Nonetheless,
almost 50 years ago Sandage \cite{sandage} tried to make
(\ref{mcvittie}) an observational tool but concluded that this
was not possible given the technology available to him. About
30 years ago Davis and May \cite {davisandmay}, with
appropriate caution, suggested that direct measurements of
$\dot{z}$ might be feasible on timescales of decades.  There have
been arguments that peculiar accelerations would invalidate
attempts to measure (\ref{mcvittie}) \cite{phillipps} countered by
arguments that in a suitable sample this would not be the case
\cite{lake1}. The latter argument seems to have survived and it
was Loeb \cite{loeb}, many years later, who pointed out that the
Ly$\alpha$ absorption-line forest on the continuum flux from
quasars could provide an ideal template for the observation of
$\dot{z}$. Recently, Corasaniti, Huterer and Melchiorri \cite
{chm} have reviewed some of the many ambitious proposals afoot for
new generations of very large telescopes along with spectrographs
of unprecedented resolution \cite{clar}. In particular, the CODEX
(Cosmic Dynamics Experiment) spectrograph should be able to detect
$\dot{z}$ in distant quasi-stellar objects ($z\geq2$) over a single decade \cite{codex}.
It would seem that observations of $\dot{z}$ might, in the relatively short term,
finally be within our grasp. However, the usefulness of such observations would, as we show here, not be
restricted to high redshifts. Low redshift ($z<2$) Ly$\alpha$ systems (which requires UV space-based systems) have already produced results \cite{penton} but much improvement will be needed to actually measure the low-redshift effects discussed here.

\bigskip

Assuming $H_{o}$ is known \cite{hubble}, (\ref{mcvittie}) provides
cosmological information through $H(z)$. Indeed, for any
background model, all properties of the model can be obtained by
taking higher $t$ - derivatives of $z$ \cite{lake}. However, $t$ -
derivatives of order $N$ produce factors of $H_{o}^{N}$ and so
this approach will remain of theoretical interest but of no
practical value beyond $N=1$. Throughout the present analysis
$N=1$. The approach used here is to explicitly use the redshift
information contained within $Z$ itself so that no higher powers
of $H_{o}$ enter. In this regard it is useful to start with a
comparison of the standard approach based on luminosity (or
other well-defined) cosmological distances.
\section{Luminosity Distances}
The standard approach in cosmology is to fit measurements to a
truncated Taylor series approximation to the luminosity distance
which, in general,  we can write as
\begin{equation}\label{luminosity}
d_{L}(z)=a_{o}(1+z)\mathcal{S}_{k}\left(\frac{c}{a_{o}}\int_{o}^{z}\frac{dx}{H(x)}\right)
\end{equation}
where $\mathcal{S}_{k}(x) \equiv \sin(\sqrt{k}x)/\sqrt{k}$. We
start with the expansion
\begin{equation}\label{taylor}
H(z)=H_{o}+H^{'}_{o}z+\frac{1}{2}H^{''}_{o}z^2+...
\end{equation}
where $^{'}\equiv d/dz$ (and, as usual, $H^{'}_{o}$ means
$dH/dz|_{z=0}$ etc.). It is customary to evaluate the coefficients
in the Taylor expansion in terms of $t$ - derivatives of the scale
factor evaluated at $t_{o}$. To do this consider $a_{o}$ fixed and
look back along the past null cone to write the familiar relation
\begin{equation}\label{othertdot}
\frac{dz}{d t_{e}}=-(1+z)H(z).
\end{equation}
This must not be confused with the notation used for $\dot{z}$ in
(\ref{mcvittie}). From (\ref{taylor}) and (\ref{othertdot}) we
obtain
\begin{equation}\label{taylorL}
\frac{H_{o}}{c}d_{L}(z)=\sum _{i=1} \mathcal{D}_{i}z^{i}
\end{equation}
where the $\mathcal{D}_{i}$ are dimensionless constants with
$\mathcal{D}_{1}=1$ but the $\mathcal{D}_{i}$ for $i>1$ contain
time derivatives of the scale factor ($da/dt|_{o}$) of order up to
$i$  and terms at that order of differentiation come in linearly
\cite{series}. Whereas the use of $d_{L}$ is standard, it is not
necessarily the best distance measure to use \cite{visser}. (Note,
however, that the use of other distance measures merely changes
the power of $(1+z)$ in (\ref{luminosity}) and this does not
change the fundamentals of the procedure under discussion.)
Moreover, let us note here that one need never introduce $t$
derivatives but rather one could proceed directly from
(\ref{luminosity}), or any other well-defined distance. For
$d_{L}$, for example, we find (\ref{taylorL}) with
$\mathcal{D}_{1}=1$ but now with
\begin{equation}\label{D2}
\mathcal{D}_{2}=\frac{1}{2}(1+Z^{'}_{o}),
\end{equation}
and
\begin{equation}\label{D3}
\mathcal{D}_{3}=\frac{1}{6}\left(Z^{''}_{o}-1-Z^{'}_{o}(1-2
Z^{'}_{o})-\frac{kc^2}{H_{o}^2a_{o}^2}\right)
\end{equation}
and so on, where $Z$ is defined by (\ref{mcvittie}) (and, as
usual, $Z^{'}_{o}$ means $dZ/dz|_{z=0}$ etc.). It is at this
order, $\mathcal{D}_{3}$, that a degeneracy (the last term above)
enters \cite{weinberg}. The use of $z$ derivatives can, of course,
be directly related to the more usual expansion in terms of $t$
derivatives. In particular, one easily finds
\begin{equation}\label{acceleration}
Z^{'}_{o}=-q_{o}
\end{equation}
and
\begin{equation}\label{jerk}
Z^{''}_{o}=-j_{o}+q_{0}^2
\end{equation}
where $q_{o}$ and $j_{o}$ are the standard deceleration and jerk
parameters \cite{series}. Despite what distance measure one
chooses to use, the quantities $Z^{'}_{o}, Z^{''}_{o}$ and so on,
enter by way of series approximation only. In the approach I
discuss here, based on (\ref{mcvittie}), this is not the case:
encoded in $\dot{z}$, given $H_{o}$, is $Z(z)$ and therefore
$Z^{'}$, $Z^{''}$ and so on and not just their values at $z=0$. In
a sense then what we consider here is a natural generalization of
the cosmographic approach. The use of (\ref{mcvittie}) clearly
offers tremendous theoretical advantage over $d_{L}$, or other
measures of distance, as (\ref{mcvittie}) makes no assumptions
other than the cosmological principle itself. However, we must
assume a background model given, say, by way of the Friedmann
function $H/H_{o}$, a standard procedure \cite{davis}. In this
sense the present approach differs fundamentally from the use of any
distances since that approach is kinematic. In practical terms,
however, it is necessary to assume that $Z(z)$, when such
observations become practical, will be able to be measured with
sufficient accuracy and precision in order to reliably produce
derivatives. This is the central assumption made here. Whereas one
should obviously never differentiate noisy data, quality data of
the type needed may eventually play an important role in cosmology
and the view taken here is that the associated theory is worth
exploration. In what follows we consider various cosmological
models and express the associated model parameters in terms of the
observables, $Z$ and its $z$ derivatives. (In a very recent paper,
Balbi and Quercellini \cite{bq} have also explored a variety of
interesting cosmological models. In terms of the present analysis,
their examination is restricted to a study of $Z$ alone.) It is to
be reemphasized that no series approximations are used. The
resultant formulae are, albeit superficially clumsy at times,
exact.
\section{The LCDM Model}
Consider an arbitrary number of non-interacting species
\cite{species} so that
\begin{equation}\label{species}
\left(\frac{H}{H_{o}}\right)^2=\;\sum_{i}\Omega_{io}(1+z)^{3(1+w_{i})}.
\end{equation}
If we are interested in the direct measurement of cosmological
parameters (say the $\Omega_{io}$ for the above model, assuming
the constants $w_{i}$ are given) via $\dot{z}$ alone, then we must
recognize that (\ref{mcvittie}) (given (\ref{species})) provides
but one equation when there are $n-1$ unknowns for $n$ species.
(The boundary condition on (\ref{species}) is the $z
\rightarrow 0$ limit $\sum_{i}\Omega_{io}=1$.) For example, in the
LCDM model (\ref{mcvittie}) alone can provide unambiguous
cosmological information only if the Universe is, say, assumed to
be spatially flat a priori. That is, there is no test of the
assumption of flatness available. The idea exploited here is to
explicitly use the redshift information contained within $Z$
itself so that no higher powers of $H_{o}$ enter and to exploit
this information so that we do not have to assume flatness.
Specifically, in terms of the observables $Z$ and $Z^{'}$ we find
the following exact relations within the LCDM model:
\begin{equation}\label{dust}
\Omega_{Mo}={\frac {-2\, Z^{'} z \left( 2+z \right)  \left( -Z
+1+z \right) +2\,Z   \left( z(2+z)+2-Z  \left( 1+z \right) \right)
}{{z}^{2} \left( 1+z \right)  \left( 3 + z \right)}}
\end{equation}
for the dust (including dark matter) and
\begin{equation}\label{lambda}
\Omega_{\Lambda o}={\frac {-2\,Z^{'}  z \left( 1+z \right) \left(
-Z +1+z \right) +Z  \left( -Z   ( 1 + 3\,z ) +2\, ( 1+z ) (
 1+2\,z )  \right) }{{z}^{2} \left( 3 + z \right) }}.
\end{equation}
Since $\Omega_{M o}$ and $\Omega_{\Lambda o}$ are constants,
(\ref{dust}) and (\ref{lambda}) provide internal consistency
checks on the LCDM model as we explore $Z(z)$. Let us note
that with the aide of l'H\^{o}pital's rule as $z \rightarrow 0$
(and note that we have to go to the second derivative), along with
equations (\ref{acceleration}) and (\ref{jerk}), we find that
(\ref{dust}) and (\ref{lambda}) reduce to
\begin{equation}\label{omegaq}
 \Omega_{Mo}=\frac{2(q_{o}+j_{o})}{3}
\end{equation}
and
\begin{equation}\label{omegaj}
 \Omega_{\Lambda o}=\frac{-2q_{o}+j_{o}}{3}
\end{equation}
so that $\Omega_{k o}=1-j_{o}$. Further, we have the following
internal explicit test for flatness within the $\Lambda$CDM model:
\begin{equation}\label{flatness}
Z^{'} ={\frac {Z  \left( -3\,Z
 \left( 1+z \right) ^{2}+12 z(1+z)+2(2z^3+3) \right) -
 {z}^{2} \left( 1+z \right) \left(3 +z \right) }{2\;z \left(3+z(3+z) \right)  \left( -Z +1+z \right) }} .
\end{equation}
Introducing the additional observable $Z^{''}$, we can also
perform an explicit internal check on the constancy of $\Lambda$
without the assumption of flatness. In particular, it must follow
from the observations that
\begin{equation}\label{lambdaexplicit}
Z^{''}={\frac {Z^{'} z  \left(  Z^{'}  z
  \left( 1+z \right)  \left( 3 +z \right) -2\,Z
 \left( 3+2 z(3+z) \right) +2\, \left( 1+z \right)  \left( 3+z(3+z) \right)  \right) -Z   \left( 3\, \left( 1+ z \right)
^{2} \left( 2-Z   \right) +2\,{z}^{3}
 \right) }{{z}^{2}   \left( 1+z \right) \left( 3 +z \right) \left( -Z
 +1+z \right) }}
\end{equation}
or the LCDM model fails.

\section{Variable Equation of State $w=w(z)$}

\begin{figure}[ht]
\epsfig{file=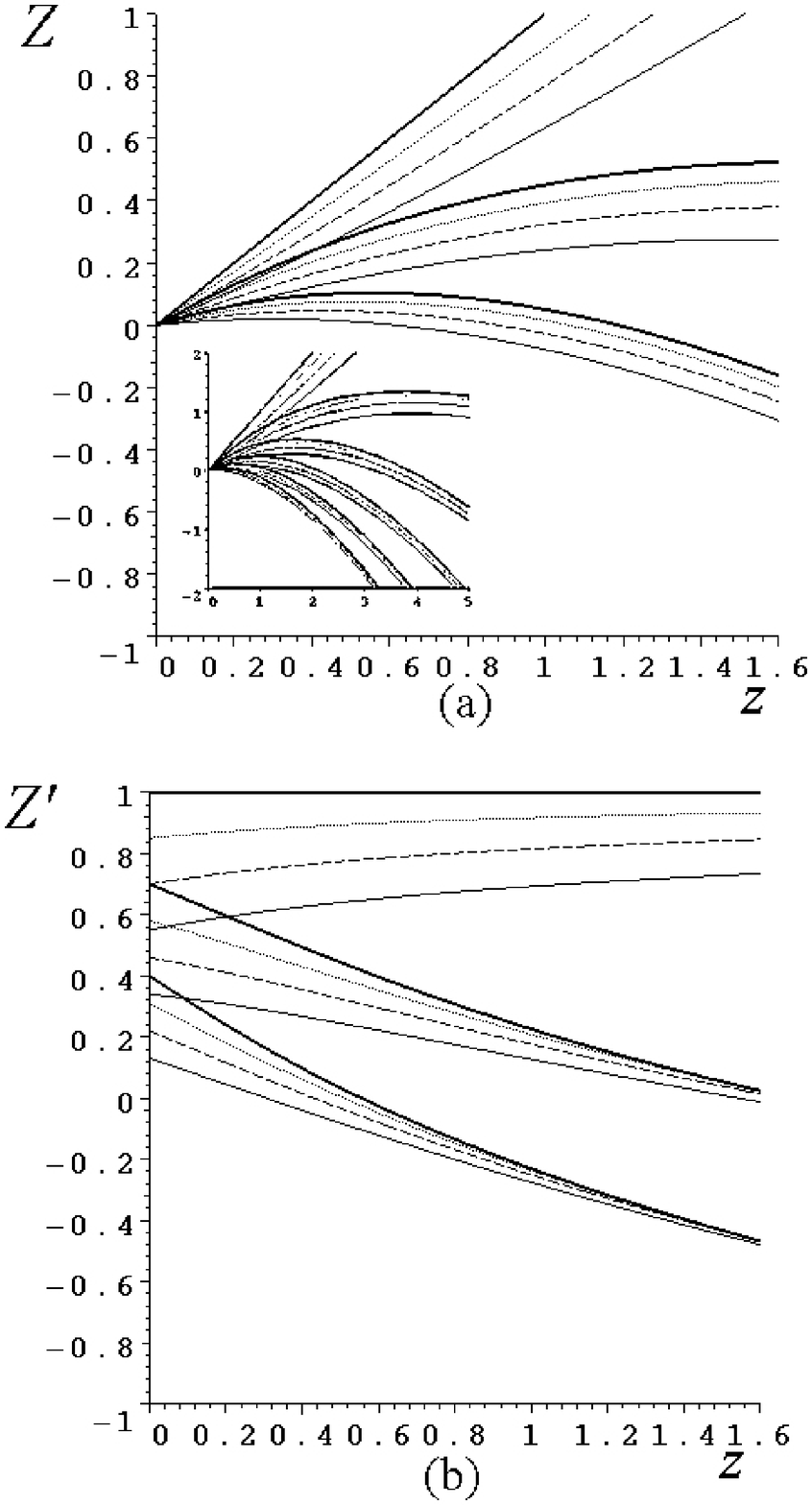,height=8in,width=4in,angle=0}
\caption{\label{Zw}The functions $Z$ (part (a)) and $Z^{'}$ (part (b)) under the
assumption of spatial flatness and constant $w$ as follows from
(\ref{zwconst}). Three values of $\Omega_{w_{o}}$ are shown;
$\Omega_{w_{o}}=1, 0.8,$ and $0.6$ and within each
four values of $w$ are shown; $w=-1$ (wide solid), $-0.9$ (dot), $-0.8$ (dash) and $-0.7$ (thin solid).
These are read top down. The insert for $Z$ also shows $\Omega_{w_{o}}=0.9, 0.7,$ and $0.5$ over a wider redshift range. The case $\Omega_{w}=-w=1$
corresponds to de Sitter space. The diagrams suggest that there
would be little advantage in trying to go to very high redshifts.
(In any event, the Ly$\alpha$ forest disappears at $z \sim 4$). }
\end{figure}

\begin{figure}[ht]
\epsfig{file=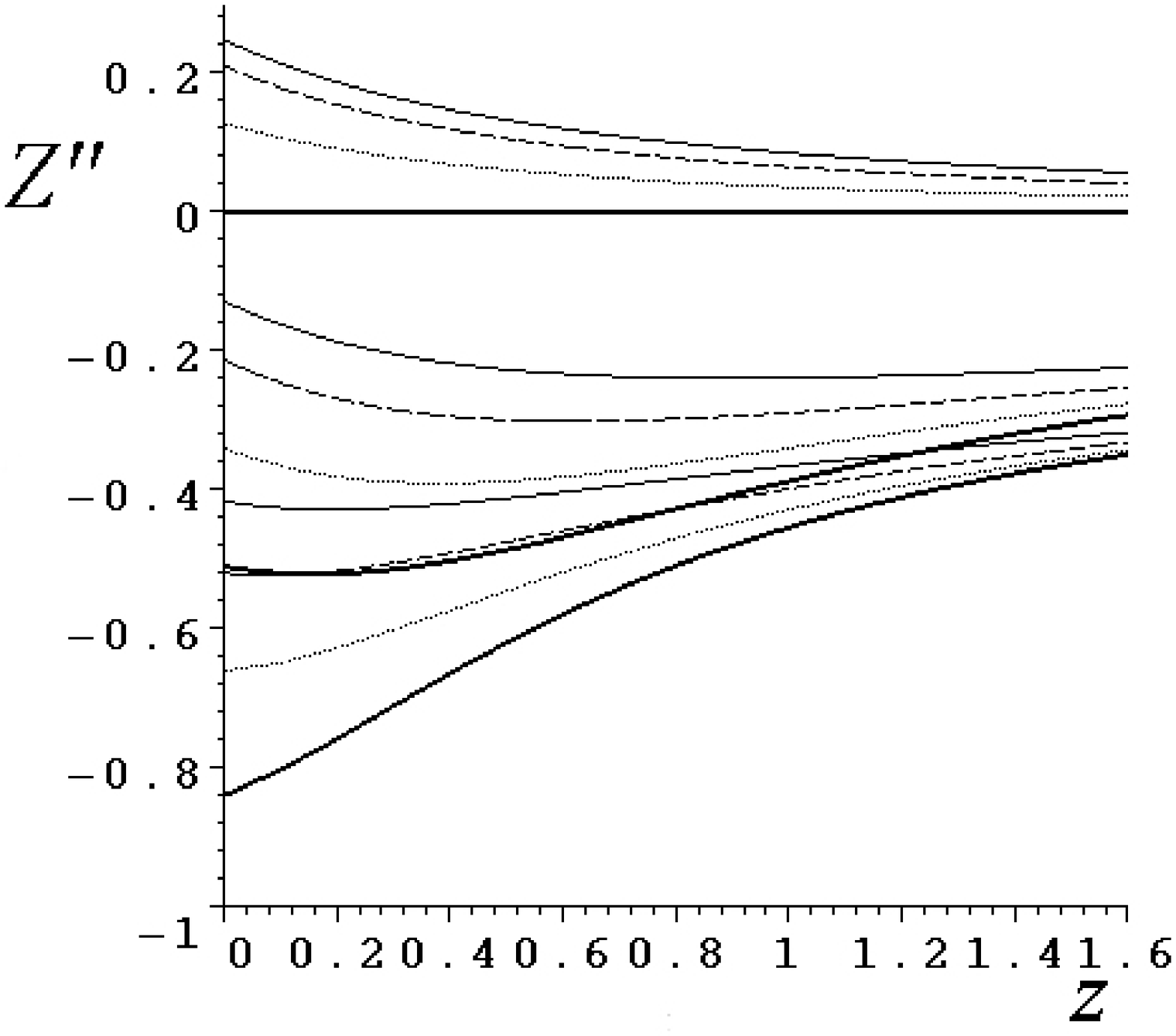,height=4in,width=4in,angle=0}
\caption{\label{Zddw}The function $Z^{''}$ under the
assumption of spatial flatness and constant $w$ as follows from
(\ref{zwconst}) with the same conditions as Figure \ref{Zw} except that the conditions on $w$ are read bottom up.}
\end{figure}

The foregoing procedure can of course be applied to any model, not
just the standard LCDM model. One popular trend today,
despite the Lovelock theorem \cite{lovelock}, is to generalize the
LCDM model by removing the assumption that
$w_{\Lambda}=-1$ and to replace it with a free (and often entirely ad hoc)
constant parameter $w$. For example, under the assumption of spatial flatness
we then have
\begin{equation}\label{zwconst}
Z=1+z-\sqrt{ \left( 1+z \right) ^{3} \left(
1-\Omega_{wo}+\Omega_{wo}
 ( 1+z) ^{3\,w} \right) }  \; ,
\end{equation}
so that
\begin{equation}\label{zdlimconstw}
Z^{'}(0)=-\frac{1}{2}-\frac{3}{2}\Omega_{{wo}}w
\end{equation}
and
\begin{equation}\label{zddlimconstw}
Z^{''}(0)=\frac{9}{4}\Omega_{{wo}}{w}^{2} ( \Omega_{{wo}}-2 ) -3\Omega_{{wo}
}w-\frac{3}{4}.
\end{equation}
With the aide of (\ref{acceleration}) and (\ref{jerk}) then
\begin{equation}\label{q0constw}
q_{0}=\frac{1}{2}+\frac{3}{2}\Omega_{{wo}}w
\end{equation}
and
\begin{equation}\label{joconstw}
j_{0}=1+\frac{9}{2}\Omega_{{wo}}w(1+w)
\end{equation}
compared with $q_{o}=1/2-3\Omega_{\Lambda o}/2$ and $j_{o}=1$ in the flat LCDM model.
The situation is summarized in Figures \ref{Zw} and \ref{Zddw}. These diagrams
suggest, for example, that $\Omega_{w_{o}}$ could be obtained from
$Z^{'}$ at high redshift and compared with $Z$ in the same
redshift range. Further, $w$ might be obtained from $Z$ at high
redshift and compared to $w$ obtained from $Z^{'}$ and $Z^{''}$ at
low redshift to check the assumption $w^{'}=0$. Although such a
comparison would require very different technologies, this comparison serves to point out that
the usefulness of $\dot{z}$ observations would not be restricted
to high redshifts.

\bigskip

If we assume spatial flatness and continue to use $Z, Z^{'}$ and
$Z^{''}$ as obervables then we need not assume $w$ is constant.
Rather, we can actually solve for $w(z)$. In the usual way we now
obtain
\begin{equation}\label{variablew}
\frac{H}{H_{o}}=\sqrt{ \left( 1+z \right) ^{3} \left(
\Omega_{Mo}+\Theta(z) \Omega_{wo} \right)},
\end{equation}
where
\begin{equation}\label{delta}
\Theta(z) \equiv \exp (3 \int_{0}^{z}\frac{w(x)}{1+x}dx).
\end{equation}
In this case writing $w=w(z)$ we find
\begin{equation}\label{mvar}  \Omega_{Mo}={\frac {
\left( -Z  +1+z \right)  \left( 2\,Z^{'} \left( 1+z \right) -3\,Z
 \left( 1+w   \right) + \left( 1+z \right)  \left( 3\,
w  +1 \right)  \right) }{3w  \left( 1+ z \right) ^{3}}}
\end{equation}
and
\begin{equation}\label{wvar}
\Omega_{wo}={\frac {-2\,Z^{'}  \left( 1+z \right)
 \left( -Z +1+z \right) +Z   \left( -
3\,Z   \left( 1+w   \right) +2\,
 \left( 1+z \right)  \left( 3\,w  +2 \right)  \right)
+ \left( 1+z \right) ^{2} \left( 3\,zw  -1 \right) }{3w
  \left( 1+z \right) ^{3}}}
\end{equation}
where now $w(z)$ follows from
\begin{equation} \label{wdot}
w \left( z \right) =\frac{{e^{\int \!-{\frac {f \left( z \right)
}{ \left( 1 +z \right) h \left( z \right) }}{dz}}}}{\int
\!3\,{e^{\int \!-{ \frac {f \left( z \right) }{ \left( 1+z \right)
h \left( z \right) }}{ dz}}} \left( 1+z \right) ^{-1}{dz}+{\it C}
}
\end{equation}
where ${\it C}$ is a constant (provided by the theory that gives
$w(z)$) and the observables $h(z)$ and $f(z)$ are given by
\begin{equation}\label{h}
 h(z) \equiv  \left( -Z  +1+z \right)  \left( -3\,Z + \left( 1+z \right)  ( 1+2\,Z^{'}
)  \right)
\end{equation}
and
\begin{eqnarray}\label{f}
f(z) \equiv -2\,Z^{''}  \left( 1+z \right) ^{2} \left( -Z
 +1+z \right) -2\, \left( 1+z \right) Z^{'}  \left( - ( Z^{'} +3
)  \left( 1+z \right) +5\,Z  \right) \\ \nonumber +Z
  \left(9\,Z -8(1+z) \right) +
 \left( 1+z \right) ^{2}.
\end{eqnarray}
Note that $\Theta$ does not enter. It should be clear that the
introduction of $Z^{'''}$ as an observable would allow us to map
out $w(z)$ without the assumption of spatial flatness, but the
associated algebra is not repeated here.

\section{Recovering The Flat LCDM Model}

As a check of the above let us note that if we use
(\ref{flatness}) in (\ref{dust}) and (\ref{lambda}), or $w(z)=-1$
and (\ref{flatness}) in (\ref{mvar}) and (\ref{wvar}), we obtain
the following simple forms for the flat $\Lambda$CDM model:
\begin{equation}\label{specialm}
\Omega_{Mo}={\frac { \left( -Z +2+z \right) \left(-Z +z\right) }{
z \left( 3+z(3+z)\right)}}
\end{equation}
and
\begin{equation}\label{speciall}
\Omega_{\Lambda o}={\frac {Z   \left( -Z
 +2(1+z) \right) +z \left( 1+z \right) ^{2}}{ z\left( 3+z(3+z) \right)
 }}
\end{equation}
 so that now the $z\rightarrow 0$ limit gives us the familiar relations (\ref{omegaq}) and (\ref{omegaj})
 with $j_{0}=1$. Similarly, substituting from (\ref{flatness}) into
(\ref{lambdaexplicit}), or $w(z)=-1$ into the derivative of
(\ref{wdot}) and using (\ref{flatness}), we obtain
\begin{equation}\label{z''simple}
Z^{''}={\frac {3 \left( 1+z \right)  \left( -Z +z
 \right)  \left( -Z +2+z \right)  \left( Z  \left( 3(\,z-1)+{z}^{2} \left( 3 +z\right)  \right)  \left( Z -2(1+z) \right) +z \left( 6+z \left( 3+z \right)
 \right)  \left( 1+z \right) ^{2} \right) }{4 {z}^{2} \left( Z  -1-z \right) ^{3} \left( 3(1+\,z)+{z}^{2} \right) ^{2}}}
\end{equation}
which, in addition to a check on the foregoing, reiterates the
fact that the only observable needed in the flat LCDM case
is $Z$ alone.

\section{Acceleration Without $\Lambda$}

\begin{figure}[ht]
\epsfig{file=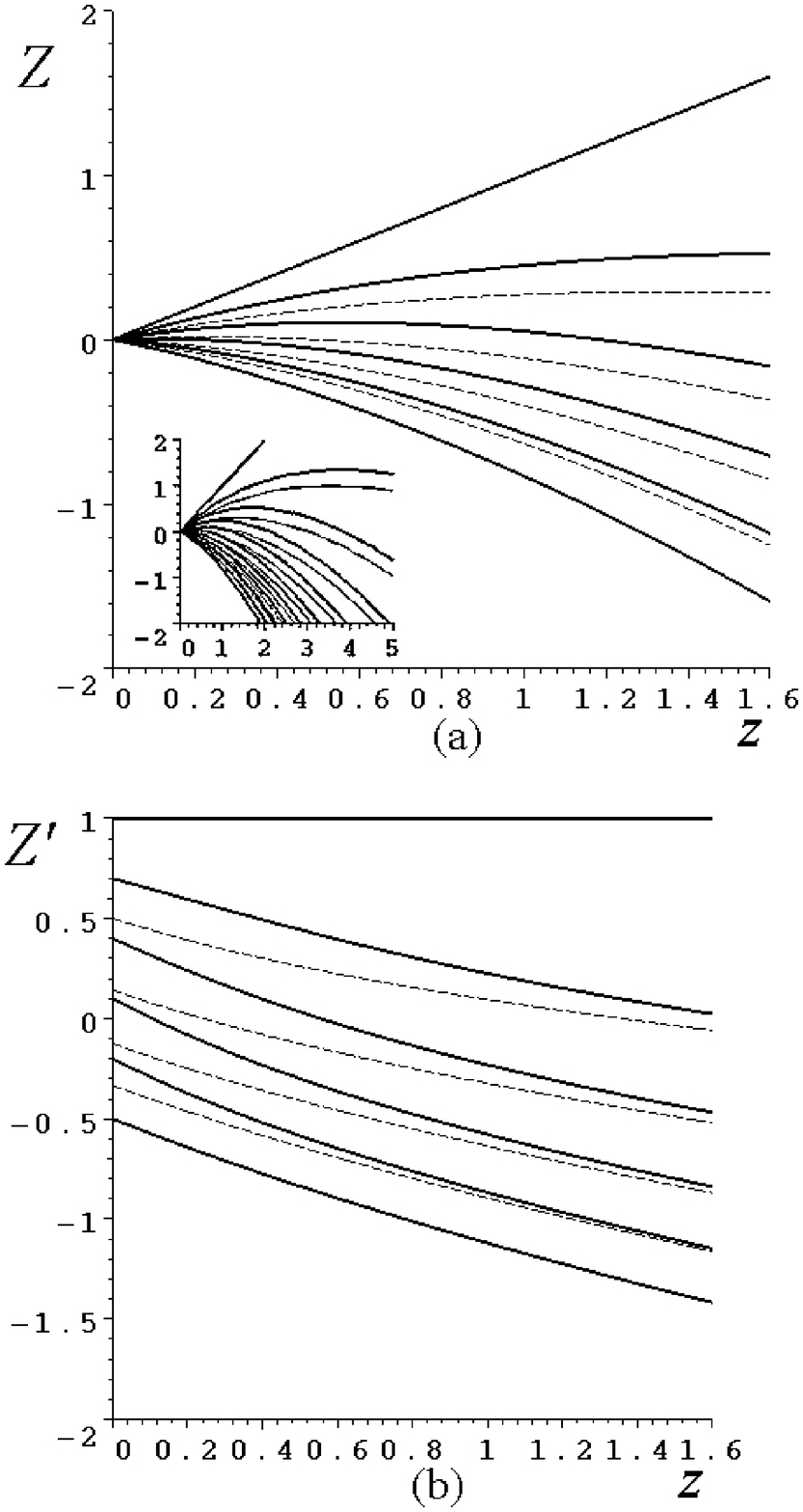,height=8in,width=4in,angle=0}
\caption{\label{lcdmb}The functions $Z$ (part (a)) and $Z^{'}$ (part (b)) as follows from
(\ref{flatbranez}) and (\ref{specialz}). The LCDM model
corresponds to the solid lines. In all cases $\Omega_{Mo}$ varies
from $1$ (bottom coincident curve, that is the Einstein-de Sitter
universe) to $0$ (top coincident curve, that is de Sitter space)
in intervals of $0.2$. The insert for $Z$ also shows more values of $\Omega_{Mo}$ over a wider redshift range.}
\end{figure}

\begin{figure}[ht]
\epsfig{file=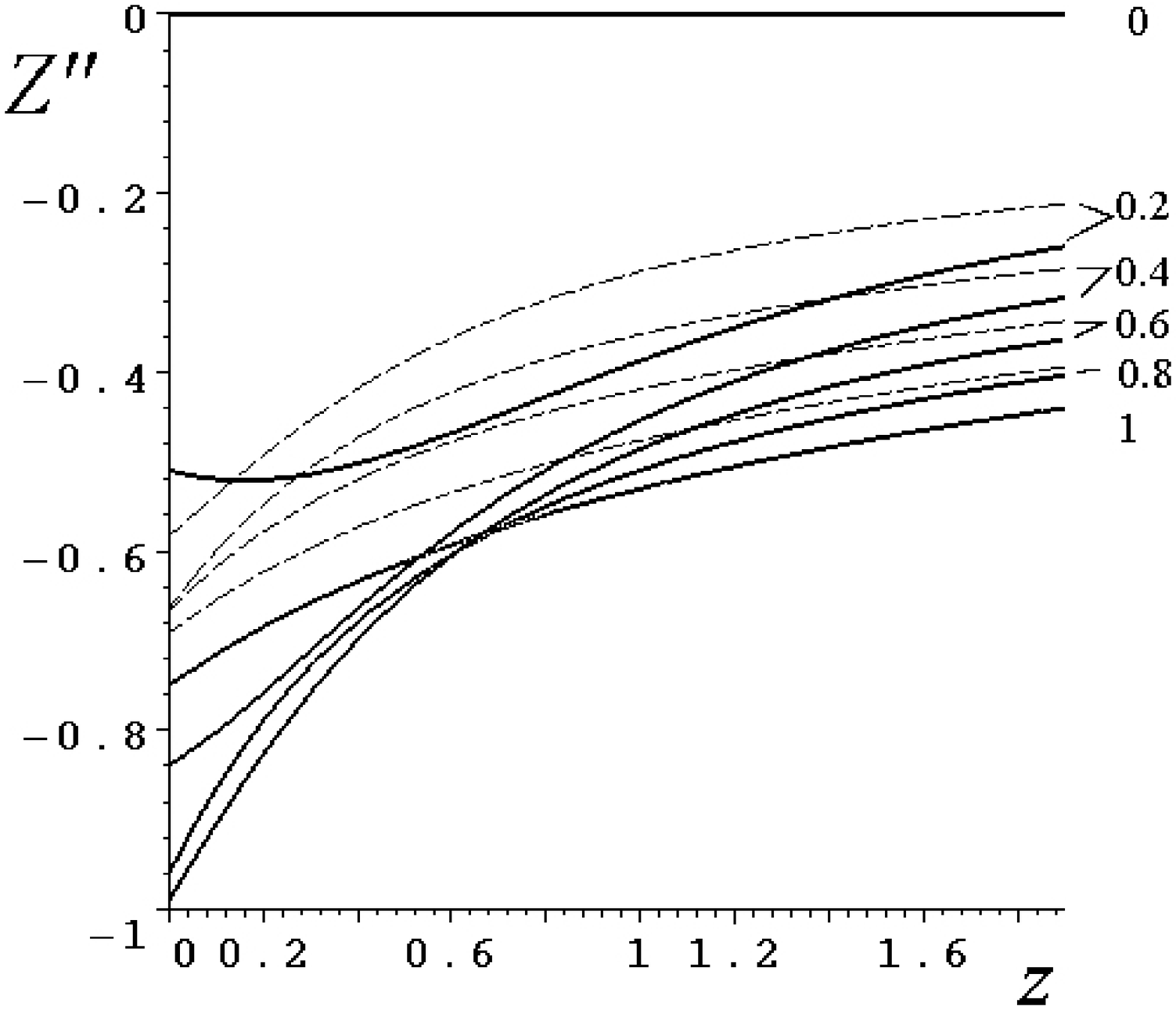,height=4in,width=4in,angle=0}
\caption{\label{lcddmbc}The function $Z^{''}$ from
(\ref{flatbranez}) and (\ref{specialz}) with the same conditions as Figure \ref{lcdmb}. The values of $\Omega_{Mo}$ are shown on the right.}
\end{figure}

The idea that the accelerated Universe could be the result of
extra dimensions - for example, the so-called DGP models \cite{dgp}, is widely
discussed in the literature. Cosmological tests based on $d_{L}$
in these models have been studied by Deffayet, Dvali and Gabadadze
\cite{deffayet} and more recently by Maartens and Majerotto
\cite{maartens}. Here we consider the spatially flat case so that
we need only consider $Z$. In this model we have
\begin{equation}\label{flatbraneh}
\frac{H}{H_{o}}=\frac{(1-\Omega_{Mo})}{2}+\sqrt{\left(\frac{(1-\Omega_{Mo})}{2}\right)^2+\Omega_{Mo}(1+z)^3}
\end{equation}
so that subject to the boundary condition $Z(0)=0$ we have
\begin{equation}\label{flatbranez}
Z=\frac{\Omega_{Mo}+1}{2}+z-\sqrt { \left( \frac{\Omega_{Mo}+1}{2}
\right) ^{2 }+\Omega_{Mo}z \left( 3+3\,z+{z}^{2} \right) }
\end{equation}
so that
\begin{equation}\label{zdbranew}
Z^{'}(0)=\frac{1-2 \Omega_{Mo}}{\Omega_{Mo}+1}
\end{equation}
and
\begin{equation}\label{zddbranew}
Z^{''}(0)=\frac{6\Omega_{Mo}(3\Omega_{Mo}-(\Omega_{Mo}+1)^2)}{(\Omega_{Mo}+1)^3}.
\end{equation}
Again with the aide of (\ref{acceleration}) and (\ref{jerk}) we now have
\begin{equation}\label{q0brane}
q_{o}={\frac {2\,\Omega_{{Mo}}-1}{\Omega_{{Mo}}+1}}
\end{equation}
and
\begin{equation}\label{jobrane}
j_{o}={\frac {10\,{\Omega_{{Mo}}}^{3}+3\,\Omega_{{Mo}}+1-6\,{\Omega_{{Mo}}}^{2}
}{ \left( \Omega_{{Mo}}+1 \right) ^{3}}}.
\end{equation}
Relation (\ref{flatbranez}) can be compared with the spatially flat LCDM model
where from (\ref{specialm}) we obtain
\begin{equation}\label{specialz}
 Z=z+1-\sqrt {1+\Omega_{{Mo}}z \left( 3+3\,z+{z}^{2} \right) }
\end{equation}
so that $q_{o}=3\Omega_{Mo}/2-1$ and of course $j_{o}=1$.
A comparison is carried out in Figures \ref{lcdmb} and \ref{lcddmbc}. The most
noticeable feature is perhaps the fact that for the same
$\Omega_{Mo}$ the models predict rather different values for
$q_{o}$ and $j_{o}$.This again serves to point out that the
usefulness of $\dot{z}$ observations would not be restricted to
high redshifts.

\section{Interacting dark energy}
\begin{figure}[ht]
\epsfig{file=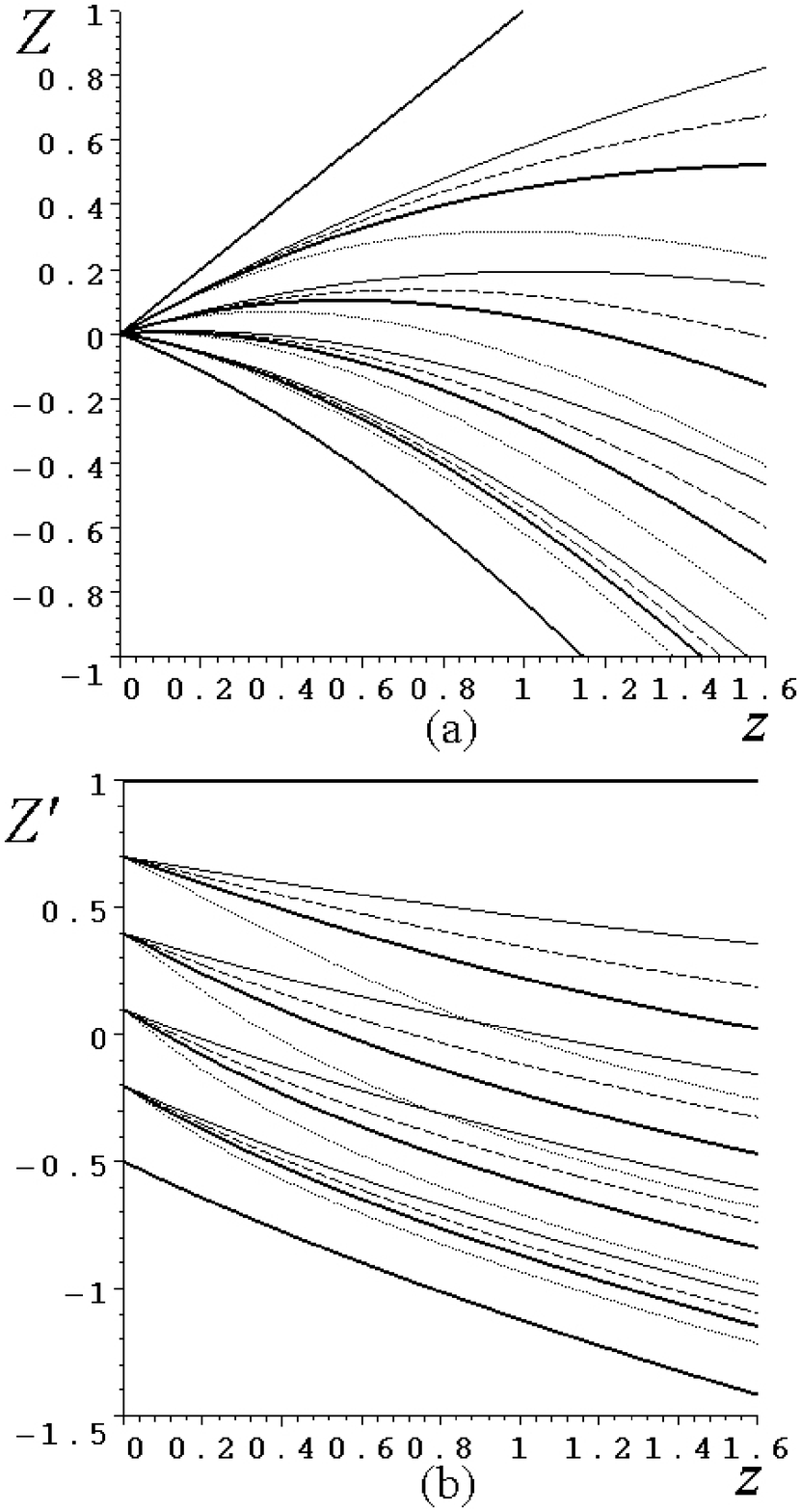,height=8in,width=4in,angle=0}
\caption{\label{lcdint}The functions $Z$ (part (a)) and $Z^{'}$ (part (b)) as follows from
(\ref{interactingflat}). In all cases $\Omega_{Mo}$ varies
from $1$ (bottom coincident curve, that is the Einstein-de Sitter
universe) to $0$ (top coincident curve, that is de Sitter space)
in intervals of $0.2$. Within each
value of $\Omega_{Mo}$ four values of $\delta$ are shown; $\delta=0.75$ (dot), $1$ (thick solid = LCDM), $1.2$ (dash) and $1.5$ (thin solid).}
\end{figure}

\begin{figure}[ht]
\epsfig{file=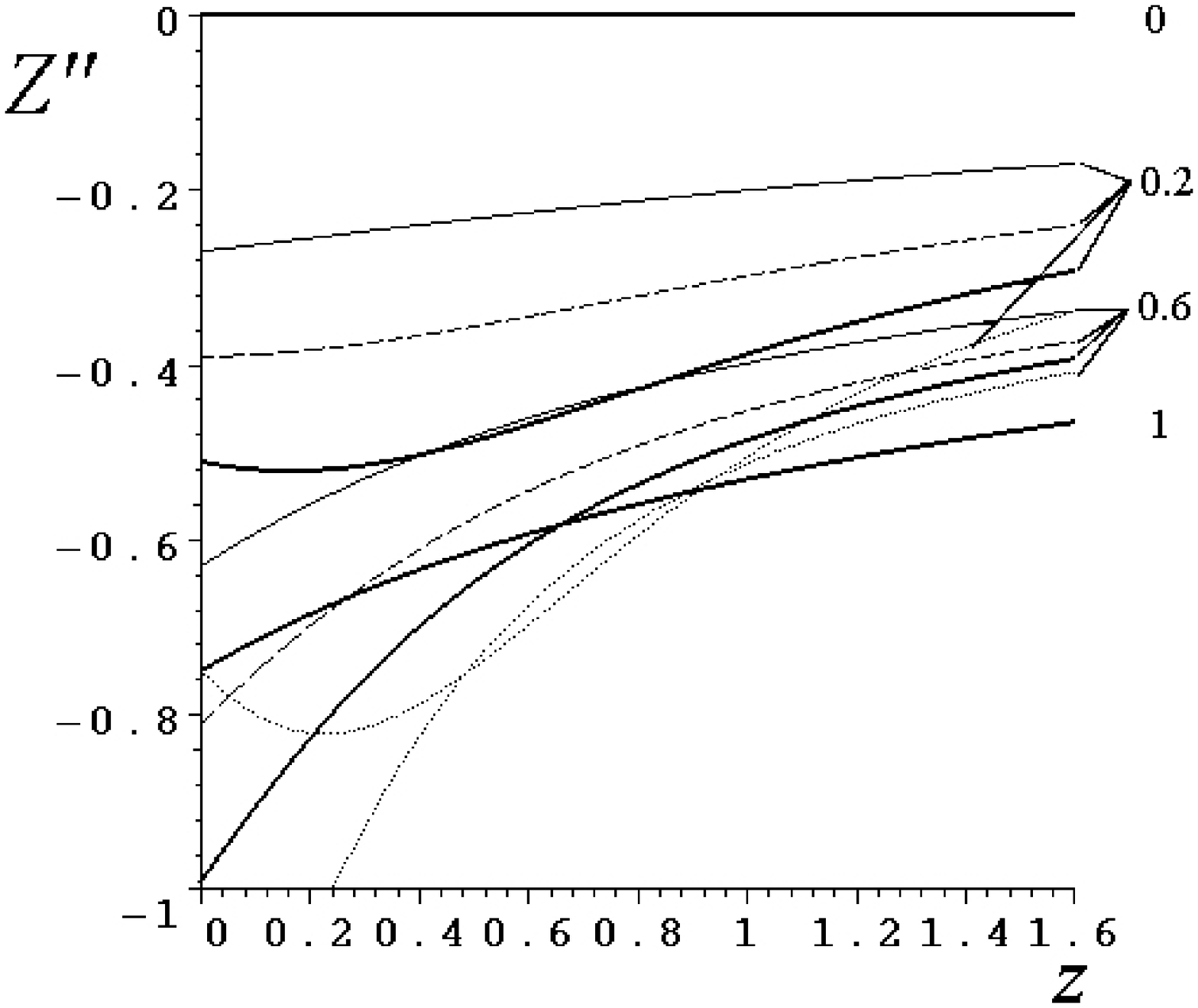,height=4in,width=4in,angle=0}
\caption{\label{lcddint}The function $Z^{''}$ from
(\ref{interactingflat}) with the same conditions as Figure \ref{lcdint}. The values of $\Omega_{Mo}$ are shown on the right.}
\end{figure}

It is natural to consider the coupling between dark energy and matter and there are many explicit coupling procedures considered in the literature. Here we use the parametrization of Majerotto, Sapone and Amendola \cite{majerotto} to write
\begin{equation}\label{interactingh}
    \left(\frac{H}{H_{\;o}}\right)^2=\Omega_{k}(1+z)^2+(1+z)^3
    (1-\Omega_{k})\left(1-\frac{\Omega_{de}}{1-\Omega_{k}}
    (1-(\frac{1}{1+z})^{\xi})\right)^{\frac{-3w}{\xi}}
\end{equation}
and setting $w=-1$, $\delta=3/\xi$ and considering the spatially flat case we have
\begin{equation}\label{interactingflat}
    Z=1+z-\sqrt{(1+z)^3\left(\Omega_{Mo}+
    (1-\Omega_{Mo})\left(\frac{1}{1+z}\right)^{\frac{3}{\delta}}\right)^{\delta}}
\end{equation}
so that with $\delta=1$ we recover the LCDM model and (\ref{interactingflat}) reduces to (\ref{specialz}).
We now have
\begin{equation}\label{zdint}
Z^{'}(0)=1-\frac{3 \Omega_{Mo}}{2},
\end{equation}
exactly as in the LCDM model, and
\begin{equation}\label{zddint}
Z^{''}(0)=\frac{3\Omega_{Mo}(6(\Omega_{Mo}-1)+2\delta(1-\frac{3\Omega_{Mo}}{2}))}{4 \delta}
.
\end{equation}
Again with the aide of (\ref{acceleration}) and (\ref{jerk}) we now have
\begin{equation}\label{q0int}
q_{o}=\frac{3 \Omega_{Mo}}{2}-1
\end{equation}
and
\begin{equation}\label{joint}
j_{o}={\frac {9\,{\Omega_{Mo}}^{2} \left( \delta-1 \right) -9\,\Omega_
{Mo} \left( \delta-1 \right) +2\,\delta}{2\delta}}
.
\end{equation}
A comparison is carried out in Figures \ref{lcdint} and \ref{lcddint}.
\section{Discussion}
Measurements of $\dot{z}$ of sufficient quality,
without the introduction of further time derivatives, would not
only allow a detailed verification of the LCDM model with
a minimum number of assumptions, but also a mapping of the dark
energy equation of state $w(z)$ should the LCDM model
fail. The procedure given here can be immediately applied to any model for which
$H(z)/H_{o}$ can be written out explicitly. A consideration
of brane-world scenarios and interacting dark energy models serves to emphasize the
fact that the usefulness of such observations would not be
restricted to high redshifts.

\begin{acknowledgments}
This work was supported by a grant from the Natural Sciences and
Engineering Research Council of Canada.
\end{acknowledgments}


\begin{thebibliography}{}
\bibitem{notation}We use a standard Robertson-Walker background
with scale factor $a(t)$, where $t$ is the proper time of comoving
observers, and write $H \equiv \dot{a}/a$ where $^{.} \equiv
d/dt$. The current epoch is designated by a subscript $o$ and a
subscript $e$ stands for the emitter. We usually write $a(t_{o})
\equiv a_{o}$. Our choice of gauge (which of course does not
affect any obeservables) is $k=\pm1,0$ for the spatial curvature
(and not $a_{o}=1$) where $[k]=L^{-2}$.
\bibitem{detect}By the detection of $\dot{z}$ we of course mean that over
the time interval $\Delta t$ the observed $z$ changes from say
$z_{1}$ to $z_{2}$ where
\begin{eqnarray}
H_{o} \Delta t =\int_{z_{1}}^{z_{2}}\frac{d z}{Z} \approx
\frac{\Delta z}{Z}.
\end{eqnarray}
\bibitem{hubble}Throughout I assume that a reliable value for
$H_{o}$ is known (and not determined from $\dot{z}$).
\end{thebibliography}
\end{document}